\documentstyle{article}

\title{\bf Jet production from the perturbative QCD pomeron with a
running coupling constant}
\author{Mikhail Braun $^{a,b)}$\thanks{
 Permanent address: Dep. High-Energy Physics, University of St. Petersburg,
198904 St.Petersburg, Russia}
, Gian Paolo Vacca  $^{b)}$ \\
$^{a)}$Department of Particles, University of Santiago de Compostela\\
$^{b)}$Department of Physics, University of Bologna\\
$^{b)}$Istituto Nazionale di Fisica Nucleare - Sezione di Bologna.}
\def\beq{\begin{equation}}
\def\eeq{\end{equation}}
\def\bea{\begin{eqnarray}}
\def\eea{\end{eqnarray}}
\def\noi{\noindent}
\def\kt{k_{\bot}}
\def\ku{k_1}
\def\kup{k'_1}
\def\qu{q_1}
\def\qup{q'_1}
\def\oq{\omega(q)}
\def\oa{\omega(q_{1})}
\def\ob{\omega(q_{2})}
\def\eq{\eta (q)}
\def\ea{\eta (q_{1})}
\def\eb{\eta (q_{2})}
\def\ec{\eta (q'_{1})}
\def\ed{\eta (q'_{2})}

\def\lb{\label}
\begin{document}
\maketitle
\medskip
\centerline{\bf Abstract.}

An analysis of minijet production from the hard pomeron with a running coupling
constant is performed.
Two supercritical pomerons found in the numerical study are taken into
account. The calculated inclusive jet production rate is finite at
small $k_{\bot}$ and behaves like $1/k^{-4}_{\bot}$ at high $k_{\bot}$
modulo factors coming from logarithmic terms. The average $k_{\bot}$
is found to be very large ($\sim 10-13\; GeV/c$) and practically
independent of energy. This is interpreted as an indication that 
at present energies we are still far from the asymptotics and
that, apart from supercritical pomerons, other states contribute
significantly.

\newpage
\section{ Introduction.}
In this paper we continue the analysis of a model where a running
coupling constant is included into the dynamics of reggeized gluons.
This model [1,2] is based on the bootstrap relation
\cite{lip1} between the reggeized gluon trajectory $\omega(q)$ and the
interaction kernel for a gluon pair $K_q(\qu,\qup)$,  which guarantees that
the production amplitudes satisfy unitarity in the one-reggeized-gluon-exchange
approximation  \cite{bart1}.
To satisfy it both $\omega$ and $K$ are written as functionals of a
single function $\eta(q)$ which is proportional to $q^2$ in the BFKL fixed
coupling case \cite{bfkl}. For the running coupling case the asymptotic
 form $\eta(q) \approx q^2/2\alpha_s(q^2)$ at $q \to \infty$ can be derived by
comparing the evolution equation with the GLAP equation in this 
limit [1,2].

We recall the basic equation for the pomeron with a running coupling:
\beq
(-\oa-\ob)\phi(q_{1})+\int\frac{d^{2}\qup}{(2\pi)^{2}}
K_{q}^{(vac)}(q_{1},q'_{1})\phi(q'_{1})=
E(q)\phi(q_{1})
 \lb{vac}
\eeq
where
\bea
\oq&=&-\alpha_{s}N_{c}\eq\int
\frac{d^{2}q_{1}}{(2\pi)^{2}}\frac{1}{\ea\eb},\ \
q=q_{1}+q_{2} \nonumber \\
K_{q}(q_{1},q'_{1})&=&-2T_{1}T_{2}\alpha_{s}
\left((\frac{\ea}{\ec}+\frac{\eb}{\ed})
\frac{1}{\eta(q_{1}-q_{1}')}-\frac{\eq}{\ec\ed}\right)
\lb{eigen} \nonumber
\eea
$T_{1(2)}$ are the colour operators for the two interacting gluons and
$N_{c}$  is the number of colours; in the vacuum channel 
$ T_{1}T_{2}=-N_{c} $. 
Function $\phi$ is the  amputated wave function. We will be interested in its
computation for the forward scattering  ($q=0$), so dividing by $\eta$
 once and twice we get the
semiamputated and the full wave functions, $\psi$ and $\Phi$ respectively. 
The bootstrap condition is

\beq
\int \frac{d^{2}q'_{1}}{(2\pi)^{2}}K_{q}^{(gluon)}(q,q_{1},q'_{1})=
\oq-\oa-\ob
\lb{boots} 
\eeq
where in the gluon channel  $T_1T_2=-N_c/2$. 

In the following, as in [1,2], we use a  parametrization 
$\eq=(b/2\pi)(q^{2}+m^{2})\ln ((q^{2}+m^{2})/\Lambda^{2})$
with $b=(1/4)(11-(2/3)N_{F})$ and $m\geq\Lambda$
and we restrict to the physical case $N_c=3$.  
We choose $\Lambda=0.2GeV$ for four acting flavours.

In  our  work \cite{bvv} we studied the
vacuum channel equation (\ref{vac}) numerically. Its
spectrum  relates to the pomeron  trajectory as 
$\alpha(q)=1-E(q) \approx 1+\Delta -\alpha' q^2$. It turned out
that two supercritical pomeron states appear as normalizable 
 solutions of (\ref{vac}). Choosing
$m=m_1=0.82GeV$, we found for them 
\beq \Delta_{0}=0.384,\ \ 
\alpha'_{0}=0.250\ (GeV/c)^{-2};\ \ \Delta_{1}=0.191,\ \ 
\alpha'_{1}=0.124\ (GeV/c)^{-2} \eeq

Their wave functions were also computed for the forward case $q=0$.

As an application, total cross sections for the 
$\gamma^{*}\gamma^{*}$ scattering, and more 
qualitatively, for the $\gamma^{*}p$ and $pp$ scattering, were 
studied in \cite{bvv} to see  the unitarization effects. It 
turned out that the contribution from multiple pomeron 
exchanges became significant only for superhigh energies 
of the order of $100 TeV$. This seems to support the idea that 
 we are still very far from the asymptopia at present 
energies, so that other states from the two-gluon spectrum 
give a large contribution.

In this paper we apply our model with the two found supercritical 
pomerons to jet production. This process has been extensively 
studied in the framework of the standard BFKL approach with a 
fixed coupling [7,8]. 
As well-known, this analysis has lead to 
some far-reaching conclusions as to the importance of 
mini-jet production at high energies and the logarithmic rise 
of the multiplicity. However in the fixed coupling approach,
various {\it ad hoc} modifications of the basic BFKL model had 
to be introduced  to cut off the spectrum at low $ \kt $ and 
also to correctly reproduce the high $ \kt $ tail of the 
spectrum.  Both these problems are naturally resolved by the 
introduction of a running coupling in our model, in which no 
new parameters appear in contrast to the fixed
coupling approach.

In section 2 we derive the basic equations and study the 
asymptotic behaviour of the jet production cross section. In 
section 3 we present our numerical results. Discussion and 
conclusions follow in section 4.  In the Appendix we study
the asymptotic form of the pomeron wave functions. 

\section{General formalism.}
In this section we consider the scattering of two highly 
virtual photons (or heavy "onia") to make the derivation  
 more rigorous. To obtain the formula for the inclusive 
jet production let us recall  the total cross section 
for one pomeron exchange \cite{braun3}:  
\beq \sigma=\frac{1}{4}\int 
d^{2}rd^{2}r'\rho_{q}(r)\rho_{p}(r') 
\int\frac{d^{2}q_{1}d^{2}q'_{1}}{(2\pi)^{4}}
G(\nu,0,q_{1},q'_{1}) \prod_{i=1,2}(1-e^{iq_{i}r})(1- 
e^{iq'_{i}r'}) \lb{sigmatot} \eeq 
Here
$ q(p) $ is the momentum of the projectile (target);
$ \nu=qp=(1/2)s $; 
$G(\nu,q,q_{1},q'_{1})$ is the non-amputated pomeron Green 
function in the momentum space, $q$ being the total momentum of the
two gluons and $q_1\,(q'_1)$ being the initial (final) momentum of
 the first gluon;
 $\rho_{q(p)}$  
is the dipole colour density of the projectile 
(target).  For the virtual photons we take 
the colour dipole densities from \cite{nz1}.

In the high energy limit  the dominant term of the Green
function comes from the two mentioned supercritical pomeron 
states:
 \beq 
G_{P}(\nu,q,q_{1},q'_{1})=\sum_{i=0,1}\nu^{\alpha_{i}(q)-1}
\Phi_{i}(q_{1},q_{2})
\Phi_{i}^{\ast}(q'_{1},q'_{2})
\lb{green1}
\eeq
where $\alpha_{i}$ and $\Phi_{i}$ are the trajectories and wave functions of
the leading (0) and subleading (1) pomerons.

Minijets appear as intermediate gluon states in the Green 
function (\ref{green1}). They possess arbitrary 
$ \kt $ subject to condition 
$ \kt^{2}<<s $. To calculate the inclusive cross-section for 
their production we have only to split the Green function in 
(\ref{sigmatot}) as indicated in Fig. 1 and drop the integration over 
$ \kt $, i.e. to substitute
 \beq 
G(\nu,0,\qu,\qup) \Rightarrow \int \frac{d^2\ku d^2 \kup}{(2 
\pi)^4} G(\nu_1,0,\qu,\ku) V(\ku,\kup) G(\nu_2,0,\kup,\qup) 
\delta^{(2)} (\ku+\kt-\kup) \eeq where, for the running 
coupling case, \beq V(\ku,\kup)= 6 \frac{\eta(\ku) 
\eta(\kup)}{\eta(\ku-\kup)} - 3 \eta(0) \eeq
Remembering that $q_1+q_2=q'_1+q'_2=q=0$ we obtain
\bea
I(y,\kt)\equiv &&\frac{d^3\sigma}{dy d^2\kt} =
\int d^{2}rd^{2}r'\rho_{q}(r)\rho_{p}(r')
\int \frac{d^{2}q_{1}d^{2}q'_{1}}{(2\pi)^{4}} 
(1-e^{i \qu r})(1- e^{i \qup r'}) 
\int \frac{d^2\ku d^2 \kup}{(2 \pi)^4} \nonumber \\
&&\delta^{(2)} (\ku+\kt-\kup)
G(\nu_1,0,\qu,\ku) G(\nu_2,0,\kup,\qup) 
\bigl[6 \frac{\eta(\ku) \eta(\kup)}{\eta(\ku-\kup)} - 3 \eta(0) \bigr]
\lb{cross1}
\eea

 The appropriate kinematical variables  are the rapidity and 
 transverse momentum  of the observed gluon (jet), 
$y=\frac{1}{2} \ln \frac{k_+}{k_-}$ and $\kt$, respectively  
; we also have $s_i=2 \nu_i$ and $s_{1(2)}=\kt \sqrt{s} 
e^{-(+)y}$.  It is  convenient to use a mixed 
(momentum-coordinate) representation for the Green functions:  
\beq
G(\nu,0,r,k)= \int \frac{d^2\qu}{(2 \pi)^2} e^{i \qu r} G(\nu,0,\qu,k)
\eeq
Defining $\Delta\Phi_i(r)=\Phi_i(r)-\Phi_i(0)$ we can write
\beq
G(\nu,0,r,k)-G(\nu,0,0,k)=\sum_{i=0,1} \nu^{\Delta_i} \Delta\Phi_i(r)
 \Phi^{*}_i(k)
\eeq
We also introduce
\beq
R_i^{q(p)} \equiv {\langle \Delta\Phi_i \rangle}_{q(p)} =
\int d^2r \rho_{q(p)}(r) \Delta\Phi_i(r)
\eeq
In this notation and also using both the semiamputated and full wave 
functions we find \beq I(y,\kt)= \sum_{i,j=0,1} R_i^q 
R_j^p \nu_1^{\Delta_i} \nu_2^{\Delta_j} \int 
\frac{d^2\ku}{(2\pi)^4} \Bigl[6 \frac{\psi_i(\ku) 
\psi_j(\ku+\kt)} {\eta(\kt)} - 3 \eta(0) \Phi_i(\ku) 
\Phi_j(\ku+\kt) \Bigr] \lb{cross2} \eeq

The found supercritical pomerons are isotropic in the 
transverse space. As a result the inclusive cross-section 
$ I(y,\kt) $ also turns out to be isotropic. Therefore we can 
integrate over azymuthal angles in (12).
Defining for the full and semi-amputated  wave 
functions the integrated quantities \bea \hat{\Phi} 
(k_1,\kt)&=&\int_0^{2 \pi} d\alpha \Phi(k_1+\kt) \nonumber \\ 
\hat{\psi} (k_1,\kt)&=&\int_0^{2 \pi} d\alpha \eta(k_1+\kt)
 \Phi(k_1+\kt)
\label{angaver}
\eea
where $\alpha$ is the angle between $k_1$ and $\kt$, we 
finally obtain the inclusive cross section 
\beq 
\frac{d^2 \sigma}{dyd\kt^2}= \frac{3}{4} \sum_{i,j=0}^1
e^{-y (\Delta_i-\Delta_j)}
(\frac{\kt^2 s}{4})^{\frac{1}{2}(\Delta_i+\Delta_j)} 
 R_i^q R_j^p \int \frac{d k_1^2} { (2 
 \pi)^3} \Bigl[ 2 \frac{\psi_i(k_1) 
\hat{\psi}_j(k_1,\kt)}{\eta(\kt)} - \eta(0) \Phi_i(k_1)  
   \hat{\Phi}_j(k_1,\kt) \Bigr] \lb{cross3} \eeq

The results of numerical calculations of the cross-section 
(14) and also its generalization to more interesting cases of 
hadronic targets or/and projectiles will be discussed in the 
next section. In the rest of this section we shall study the 
asymptotical behaviour of the found inclusive jet production 
cross-section at very small and very large transverse momenta 
and also its 
$y$ -dependence.

As to the latter, all 
$ y $-dependence in (\ref{cross3}) comes from the factor 
$ \exp (-y(\Delta_{i}-\Delta_{j})) $ which has its origin in 
the existence of two different supercritical pomerons. 
Evidently in the limit 
$ s\to\infty $ this dependence dies out, since the relative
contribution of the subdominant pomeron becomes neglegible.
The model thus predicts an asymptotically flat 
$ y$ plateau at very high energies.

At small 
$ \kt $ the cross-section (\ref{cross3}) evidently goes down as
$ \kt^{\Delta_{1}} $, since all other factors  are finite in 
this limit. However one should remember that (\ref{green1}) gives the 
dominant contribution only while 
$ (\kt^{2}s) $ continues to be large. At too small 
$ \kt $, when the above quantity becomes finite, all other 
states from the spectrum of the two-gluon equation (\ref{vac}), 
hitherto neglected, begin to give comparable or even 
dominant contribution, so that the found $ \kt^{\Delta_{1}} $
behaviour ceases to be valid. 

To find  the asymptotic behaviour of the 
inclusive cross section for $ \kt \to \infty$  
 we need to know  the behaviour of the pomeron wave functions
in the momentum space at 
$ q\to\infty $ and in the ordinary space at 
$ r\to 0 $. 
In the Appendix we show that \bea \psi(q) && 
\mathop{\sim}_{q \to \infty} \quad \frac{1}{q^2} \bigl( \ln 
q^2 \bigr)^{\beta} \nonumber \\ \psi(r) && \mathop{\sim}_{r 
\to 0} \quad \bigl( \ln \frac{1}{r} \bigr)^{\beta+1} \eea
where $\beta = -1 -\frac{3}{b E}$ so that in the forward scattering case 
$\beta$ depends just on the intercept of the corresponding pomeron state.
Let us study now the behaviour of (\ref{cross2}) for 
$\kt \to \infty$.
Using  
$\Phi(q) = \psi(q)/\eta(q) \sim \frac{1}{q^4}\bigl( \ln q^2 \bigr)^{\beta-1}$
we find $\int d^2q \Phi(q) < \infty$, so that for the second term in the 
integrand of
(\ref{cross2}) we get
\beq
\int d^2 \ku \Phi_i(\ku) \Phi_j(\ku+\kt) 
\mathop{\longrightarrow}_{\kt^2 \to \infty} \Phi_j(\kt) \int d^2\ku \Phi_i(\ku)
\sim \frac{1}{\kt^4} \bigl( \ln \kt^2 \bigr)^{\beta_j-1}
\eeq

To analyze the first term we use the identity
\beq
\int \frac{d^2\ku}{(2 \pi)^2} \psi_i(\ku) \psi_j(\ku+\kt)=2 \pi
\int_0^{\infty} r dr J_0(\kt r) \psi_i(r) \psi_j(r)
\eeq
and the relation
\beq
\int_0^{\infty} r dr J_0(\kt r)f(r)=-\frac{1}{\kt}
\int_0^{\infty} r dr J_1(\kt r)f'(r)
\eeq
which is valid for naturally behaved f, such that 
$\Bigl[ r f(r) J_1(\kt r) \Bigr]_{r=0}^{r=\infty}=0$.

Putting $f(r) \sim \bigl( \ln \frac{1}{r} \bigr)^{\beta_i+\beta_j+2}$
we obtain the leading behaviour
\beq
\int d^2\ku\psi_i(\ku) \psi_j(\ku+\kt) \sim \frac{1}{\kt^2} 
\Bigl( \ln \kt^2 \Bigr)^{\beta_i+\beta_j+1}
\eeq
So at 
$ \kt\to\infty $ we find for the inclusive cross-section (12) 
\beq
I(y,\kt) \sim \sum_{i,j=0,1} R_i^q R_j^p 
 (\frac{\kt^2  s}{4})^{\frac{\Delta_i+\Delta_j}{2}}
   e^{-y (\Delta_i-\Delta_j)} \frac{1}{\kt^4} 
   \Bigl( \ln \kt^2 \Bigr)^{\beta_i+\beta_j}
\label{asymcross}
\eeq   
This asymptotics corresponds to the standard quark-counting 
rules behaviour ($\sim 1/\kt^{4}$), modified by a power factor 
due the pomeron energetic dependence and a logarithmic factor
coming from the pomeron wave function, i.e. from the running 
of the coupling. Note that in (\ref{asymcross}) the 
$ \kt $ and 
$ s $ dependencies are separated. As a result we find that the 
average 
$ \langle \kt \rangle $  is finite and independent of $ s $ and 
$ \langle \kt^{2} \rangle \sim s^\Delta $, since it formally
diverges for (20) and one has to restrict $k^2_{\bot}<s$.

For the the multiplicity
\beq
\langle n \rangle = \frac{1}{\sigma} \int \frac{d^3 \sigma}{dy d^2\kt}
		dy d^2\kt
\eeq
the standard asymptotical behaviour $ \langle n \rangle = a \ln s + b$
is obtained. Indeed integrating (\ref{cross2}) we get 
$\int dy d^2\kt \frac{d^3 \sigma}{dyd^2\kt}=B s^{\Delta_0}\ln s
+ C s^{\Delta_0}$.  Since the total cross-section has the form
$\sigma= A s^{\Delta_0}$ at large 
$ s $, we get the mentioned asymptotical expression for the 
multiplicity. Of course this result is valid only in the 
extreme limit 
$ s\to\infty $. At large but finite 
$ s $ the existence of two different pomerons leads to some 
additional non-trivial 
$ s $-dependence.

It is instructive to compare our asymptotic results with those 
obtained in the BFKL fixed coupling model. In the latter case 
the inclusive cross section is of course badly behaved in the 
$\kt \to 0$ limit due to scale invariance. It is also 
very different in the high $\kt$ limit.  At very large 
$ \kt $ such that 
$ \ln\kt\sim\sqrt{\ln s} $ one finds 
\beq \Bigl(\frac{d^2\sigma}{dyd\kt^2}\Bigr)_{BFKL} \quad
\mathop{\sim}_{\kt \to \infty} \quad a(y)\frac{ 
(\kt^2 s)^{\Delta}}{\kt^2} \frac{e^{- \ln^2 \kt^2/a^{2}(y)}} 
{\ln \kt^2\sqrt{\ln s}} \eeq
where $a^2 \sim (\ln s-4y^{2}/\ln s)$. Thus for  large 
$\kt$  the BFKL cross section goes down faster than any power.
Also one obtains that 
$ \langle \ln\kt\rangle\sim\sqrt{\ln s} $ so that both $ \kt $ and 
$ \kt^2 $ grow with $ s $. 

To bring these predictions in better correspondence with the
physical reality, as mentioned, various modifications of this
orthodox BFKL approach have been introduced. In particular
in [7,8] the fusion of gluons via  
 "fan" diagrams was assumed to take place at high gluonic 
densities, which was considered as a way to partially restore 
the s-channel unitarity.  Then, under some additional 
assumptions, a 
$ 1/\kt^4 $ asymptotic behaviour similar to 
(\ref{asymcross}) was found.
In our model such a behaviour naturally follows  without
introducing additional assumptions nor imposing the unitarity
restrictions.

\section{Numerical results.}
Taking the wave function evaluated numerically in \cite{bvv} we have
computed  the cross section (\ref{cross3}).

In Fig.2 we present $d^2\sigma/dyd\kt$ for the process
$\gamma^*\gamma^*$ (in 
units $c=1$).
We have chosen  the projectile photon to have virtuality  
$Q=5GeV/c$ and  the target one to have virtuality $P=1GeV/c$. 
The center of mass energy is $\sqrt{s}=540 GeV$.

Of course, processes involving hadronic targets or/and 
projectiles are much more interesting from the practical point 
of view. However these require some non-perturbative 
input for the colour densities of the colliding hadrons. 
A possible way to introduce it is evidently to convert 
Eq. (1) into an evolution equation in 
$ 1/x $  and take initial conditions for it from the 
existing experimental data. Postponing this complicated 
procedure for future studies, we use here a simpler 
approach, taking for the hadron (proton) a Gaussian 
colour density with a radius corresponding to the 
observed electromagnetic one. Such an approximation, in 
all probability, somewhat underestimates the coupling of 
the hadron to the pomeron, since the coupling of the 
latter to constituent gluons is neglected. Nevertheless, we 
hope that it gives a reasonable estimate for the 
inclusive cross-section. Using this approach we get the 
inclusive jet production cross-sections 
 for the $\gamma^*p$ and $pp$ scattering  shown in Fig.3 and 4 
respectively.

A common feature of jet production in all processes is that 
the cross-section reaches a maximum at 
$ \kt\approx 1\ GeV/c $ from which it monotonously goes down 
both for smaller and larger $ \kt $.

 In Fig. 5 the cross-section $d \sigma/dy$
integrated over 
$ \kt $ in the interval $0.5\div 20\ GeV/c$
is presented for the process 
$ pp $. The limitations in the numerical calculations of the 
 wave functions do not allow us to study higher values of 
$ \kt$, so that we cannot numerically reach the region where 
the asymptotical behaviour (\ref{asymcross}) is strictly valid.

  In Fig. 6 we show jet multiplicities as a function of 
$ s $ for the three studied processes $\gamma^*\gamma^*$,
  $\gamma^*p$ and $pp$. As one observes, their magnitude and 
behaviour prove to be quite similar. 

We also tried to estimate the average $ \langle
 \kt \rangle$.
Unfortunately, altough it exists according to (20), its value
results very sensitive to the high momentum tail of the
pomeron wave function, poorly determined from our numerical
calculations. To avoid this difficulty we chose to calculate the
average $\langle \ln \kt/\Lambda \rangle$ instead. This average at 
fixed $y$ depends on $s$ and $y$ only due to the existence
of two different pomerons, so that this dependence should die
out at large enough $s$. In fact the found average
$\langle \ln \kt/\Lambda \rangle$ turns our to be practically
independent of $y$ and very weakly dependent on $s$ in the whole
studied range of $s$ and
$y$,  rising from  3.97 at $\sqrt{s}=20\ GeV$ to
4.19 at $\sqrt{s}=20\ TeV$. These values imply a rather high average
$\kt$ rising from 10.6 $GeV/c$ at $\sqrt{s}=20\ GeV$ to
13.2 $GeV/c$ at $\sqrt{s}=20\ TeV$.

\section{Discussion}
Our calculations show that the introduction of a running 
coupling constant on the basis of the bootstrap condition 
cures all the deseases of the orthodox BFKL approach for jet 
production. At high 
$ \kt$  the cross-section becomes well-behaved and more or 
less in accordance with the expectations based on the quark 
counting rules. At small 
$ \kt$ no singularity occurs, although contributions from 
other states is expected  to dominate.

As to 
$ s$-  and $y $ -dependence, our predictions are even simpler 
than in the BFKL approach, since the  running of the coupling
converts the branch point in the complex angular momentum 
plane, corresponding to the BFKL pomeron, into poles, of which 
only two are located to the right of unity and contribute at
high energies. As a result at superhigh energies, when only 
the dominant pomeron survives, the 
$ y $-dependence completely disappears and the 
$ s $-dependence reduces to the standard 
$ s^{\Delta_{0}} $ factor. At smaller 
$ s$  some 
$ y $- and extra 
$ s $-dependence appears due to the existence of two 
supercritical pomerons.

With all these refinements, some basic predictions of the BFKL 
theory are reproduced. The cross-section for minijet 
production rises fast and saturates the total cross-section as 
$ s\to\infty $. Jet multiplicities rise logarithmically.
 
However $ \langle\kt\rangle $ turns out to depend on $s$ weakly
and its calculated value results pretty high, of the order
of 10-12 $GeV/c$. This should be contrasted to the experimentally
observed much lower values of $\langle \kt\rangle$ rising with
energy. A natural explanation of this discrepancy follows from
the conclusion made in [6] from the study of the structure
functions and total cross-sections in our model: at present 
energies the contribution from the two supercritical pomerons only
covers a part of the observed phenomena because we are still rather far
from the real asymptotics. The bulk of the contribution comes
from other states, which produce a much softer spectrum of
particles. The observed rise of the $\langle\kt\rangle$ is then related
to the  dying out of these subasymptotical states. Our prediction
is then that the rise of $\langle\kt\rangle$ should saturate at the
level of 10-12 $GeV/c$.

This circumstance has also to be taken into account when discussing
the numerical results presented in Figs. 2-6. Probably they also
illustrate predictions for considerably higher energies than the 
present ones, at which, according to the estimate made in [6],
they should account for $\sim 20$\% of the observed spectra.

To obtain predictions better suited for present energies
one should evidently take into account all states present
in the spectrum of the pomeron equation (1) and not only the
two supercritical pomerons. To realize this program an evolution
equation in $\nu$ following from (1) seems to be an appropriate tool.
As mentioned, it could also effectively
take into account the nonperturbative
effects related to the pomeron coupling to physical hadrons.
A work in this direction is now in progress.

\section{Appendix}
\subsection {The momentum space behaviour}
The asymtptotic form of the pomeron eigenvalue equation at high momentum  
is [1,2]
\beq
\ln\frac{\ln q^{2}}{\ln m^{2}} \psi(q)+\frac{am^{2}\ln m^{2}}{q^{2}\ln
q^{2}}={\tilde\epsilon}\psi(q)+\frac{1}{\pi}\int\frac{d^{2}q' \psi (q')}
{[(q-q')^{2}+m^{2}]\ln[(q-q')^{2}+m^{2}]}
\lb{eigen1}
\eeq
where we have put $E=(3/b)\tilde\epsilon$
(we consider the case $m=m_1$ for simplicity) 
Let us study the last (integral) term, which we denote as $D$. Changing the
variable $q'=|q|\kappa$ we present it in the form
\[
D=
\frac{1}{\pi}\int\frac{d^{2}\kappa \psi (|q|\kappa)}
{[(n-\kappa)^{2}+m^{2}/q^{2}]\ln[q^{2}(n-\kappa)^{2}+m^{2}]}\]\beq
=\frac{1}{\pi}\int\frac{d^{2}\kappa \psi (|q|(n+\kappa))}
{(\kappa^{2}+m^{2}/q^{2})\ln(q^{2}\kappa^{2}+m^{2})}
\eeq
with $n^{2}=1$.
In the last form it is evident that the leading terms at
$q\rightarrow\infty$
come from the integration region of small $\kappa$. Then we split the
total $\kappa$ space into two parts: $\kappa>\kappa_{0}$ and
$\kappa<\kappa_{0}$ where $\kappa_{0}$ is a small number
\beq
\kappa_{0}<<1
\lb{cond1}
\eeq
The contributions from these two parts we denote as $D_{1}$ and $D_{2}$,
respectively. Our first task is to show that $D_{2}$ cancels the kinetic
term in (\ref{eigen1}) irrespective of the asymptotics of $\psi(q)$.

With small enough $\kappa_{0}$ and for a "good enough" $\psi$ 
\beq
\psi (|q|(n+\kappa))\simeq\psi (|q|n)=\psi(q)
\lb{approx1}
\eeq
so that we can take it out of the integral over $\kappa$ in $D_{2}$.
Actually (\ref{approx1}) is our definition of a "good" function, so that we shall have
to check if this condition is indeed satisfied for the found asymptotical
$\psi(q)$. 

With (\ref{approx1}) $D_{2}$ simplifies to
\beq
D_{2}=\frac{\psi(q)}{\pi}\int_{0}^{\kappa_{0}^{2}q^{2}}\frac{d^{2}q' }
{({q'}^{2}+m^{2})\ln({q'}^{2}+m^{2})}=
\psi(q)\ln\frac{\ln(\kappa_{0}^{2}q^{2}+m^{2})}{\ln m^{2}}
\lb{d2_1}
\eeq
As $q\rightarrow\infty$ we assume that also
\beq
\kappa_{0}q\rightarrow\infty
\lb{limk0q}
\eeq
Evidently this condition fixes the manner in which $\kappa_{0}$ goes to zero
as $q$ turns large. This should be taken into account when veryfying
condition (\ref{cond1}). Then (\ref{d2_1}) gives
\beq
D_{2}=\psi(q)\ln\frac{\ln\kappa_{0}^{2}q^{2}}{\ln m^{2}}
\eeq
We have also
\[\ln\ln\kappa_{0}^{2}q^{2}=\ln(\ln q^{2}+\ln\kappa_{0}^{2})\]
But according to (\ref{limk0q})
\[\ln q^{2}>>|\ln\kappa_{0}^{2}|\]
so that we have
\[\ln\ln\kappa_{0}^{2}q^{2}=\ln\ln q^{2}+\frac{\ln\kappa_{0}^{2}}{\ln q^{2}}
+ \cdots\]
Then our final result is
\beq
D_{2}=\psi(q)\left(\ln\frac{\ln q^{2}}{\ln m^{2}}+O(1/\ln q^{2})\right)
\lb{d2_2}
\eeq
The first term  exactly cancels the kinetic energy ( first) term
in the pomeron equation. The correction term in (\ref{d2_2}) evidently is much
smaller than ${\tilde\epsilon}\psi(q)$, since the factor which multiplies
the function $\psi$ goes to zero in (\ref{d2_2}). So we can safely neglect it.

As a result, the asymptotic equation becomes
\beq
\frac{am^{2}\ln m^{2}}{q^{2}\ln
q^{2}}={\tilde\epsilon}\psi(q)+D_{1}
\lb{eigen2}
\eeq
with the term $D_{1}$ given by
\beq
D_{1}=
\frac{1}{\pi}\int\frac{d^{2}q' \psi
(q')\theta((q-q')^{2}-\kappa_{0}^{2}q^{2})}
{(q-q')^{2}\ln (q-q')^{2}} 
\lb{d1}
\eeq
where we have omitted the terms $m^{2}$ in the denominator because of 
(\ref{limk0q}).

As in [2] we pass to the function $\chi(q)$ defined by
\beq  \chi(q)=\psi(q)q^{2}\ln q^{2}\eeq
Multiplying (\ref{eigen2}) by $q^{2}\ln q^{2}$ we find an equation for $\chi$
\beq
am^{2}\ln m^{2}={\tilde\epsilon}\chi(q)+
\frac{1}{\pi}\int\frac{d^{2}q' \chi
(q')\theta((q-q')^{2}-\kappa_{0}^{2}q^{2})}{{q'}^{2}\ln {q'}^{2}}
\frac{q^{2}\ln q^{2}}{(q-q')^{2}\ln (q-q')^{2}}
\lb{eigen3}
\eeq

At this point we make a second assumption. Namely we assume that in the
integral term of (\ref{eigen3}) values $m<<q'<<q$ give the dominant 
contribution (as
typical for logarithmic integrals). Of course, this assumption is also to be
checked for the final asymptotics.
With this assumption, we can forget about the $\theta$ function and also
put the last factor  to unity in the integral term of (\ref{eigen3}). We obtain
\beq
am^{2}\ln m^{2}={\tilde\epsilon}\chi(q)+
\int_{0}^{q^{2}}\frac{d{q'}^{2} \chi
(q')}{{q'}^{2}\ln {q'}^{2}}
 \eeq
Differentiating with respect to $q^{2}$
\beq
{\tilde\epsilon}\frac{d\chi}{dq^{2}}=-\frac{\chi}{q^{2}\ln q^{2}}
\eeq
or
\beq
{\tilde\epsilon}\frac{d\chi}{d\ln\ln q^{2}}=-\chi
\eeq
with a solution
\beq
\chi(q)=A\exp\left(-\frac{\ln\ln q^{2}}{\tilde\epsilon}\right)
=A(\ln q^{2})^{-1/
\tilde\epsilon}\eeq
The initial function $\psi(q)$ has the asymptotics
\beq
\psi(q)=\frac{A}{q^{2}}(\ln q^{2})^{\beta}
\lb{asymptmom}
\eeq
where
\beq
\beta=-1-\frac{1}{\tilde\epsilon}=-1-\frac{3}{bE}
\eeq

Now we have to check that our assumptions are indeed fulfilled for the
found asymptotics. 

Let us begin with the second assumption that the values $m<<q'<<q$ give
the bulk of the contribution to the integral in (\ref{eigen3}). 
Evidently, in order
that the integral be dominated by large values of $q'$, it
 should diverge as $q\rightarrow\infty$. This
leads to the condition
\beq
E<0,\ \beta>-1
\eeq
So our asymptotics can only be valid for negative energies, that is, for
bound states.

Now for the second part of this assumption. To prove that values $q'<<q$
dominate we shall calculate the contribution from the region $q'>>q$ and
show that it is smaller. The corresponding integral is
\beq
I=Aq^{2}\ln q^{2}\int_{q^{2}}^{\infty}d{q'}^{2}(\ln {q'}^{2})^
{\beta-1}/{q'}^{4}
\eeq
 In terms of $x=\ln {q'}^{2}$
\beq
I=q^{2}\ln q^{2}\int_{\ln q^{2}}^{\infty}dxx^{\beta-1}\exp (-x)
\eeq
The integral over $x$ can be developed in an asymptotic series in $1/\ln
q^{2}$:
\beq
\int_{\ln q^{2}}^{\infty}dxx^{\beta-1}\exp (-x)=
-\int_{\ln q^{2}}^{\infty}x^{\beta-1}d\exp (-x)=
\frac{(\ln q^{2})^{\beta-1}}{q^{2}}+
(\beta-1)\int_{\ln q^{2}}^{\infty}dxx^{\beta-2}\exp (-x)=...
\eeq
>From this we conclude that the integral $I$ has the asymptotics
\beq
I=A(\ln q^{2})^{\beta}
\eeq
to be compared to the contribution from the region $q'<<q$ which behaves as
$(\ln q^{2})^{\beta+1}$.   We see that we have lost one power of
$\ln q^{2}$, so that the region $q'>>q$ indeed can be neglected.

Now to the assumption (\ref{approx1}). We have explicitly
\[ \psi(|q|(n+\kappa))=(A/q^{2})(n+\kappa)^{-2}(\ln q^{2}(n+\kappa)^{2})^
{\beta}=\]
\beq
(A/q^{2})(1-2n\kappa-\kappa^{2}+4(n\kappa)^{2})
(\ln q^{2})^{\beta}(1+(\beta (2n\kappa+\kappa^{2})/\ln q^{2})
\lb{approx1_2}
\eeq
and it is evident that (\ref{d1}) is satisfied with the behaviour of $\kappa$ as
indicated in (\ref{limk0q}).
Indeed take $\kappa\sim q^{-\delta}$ with $\delta<1$.
Then $q\kappa\sim q^{1-\delta}\rightarrow\infty$ and all the correcting
terms in (\ref{approx1_2}) have the order $q^{-\delta}$.

Note that this does not mean that (\ref{approx1}) is quite obvious.  It is not
valid for, say, the exponential function. In fact we have
\[\exp(aq^{2}(n+\kappa)^{2})=\exp(aq^{2})\exp
(aq^{2}(2n\kappa+\kappa^{2}))\]
and since $q\kappa$ is large the second factor cannot be neglected.

In conclusion, we have verified that our assumtions are fulfilled and
therefore the asymptotics (\ref{asymptmom}) is correct for negative energies.

\subsection{The coordinate space  behaviour}

We are also interested in the coordinate space behaviour of the semiamputated
wave function
\beq
\psi(r)= \int_0^{\infty} \frac{q dq}{2 \pi} J_0(qr) \psi(q)
\eeq

To estimate $\psi(r)$ for $r \to 0$ we make use of the asymptotic momentum
behaviour found previously (\ref{asymptmom}), so we can write
\beq
\psi(r) \approx \int_0^{q_0} \frac{q dq}{2 \pi} J_0(qr) \psi(q) +
         \int_{q_0}^{\infty} \frac{q dq}{2 \pi} J_0(qr) 
         \frac{ln(q^2)^{\beta}}{q^2}
\eeq         

The first integral for $r \to 0$ is finite; 
on the other hand the second
integral results to be not bounded. Infact using the integration variable
$y=qr$ we split the $y$ integration region in two parts, $q_0 r < y < y_0 \ll 1$
and $y \ge y_0$; defining the two contribution $I_1$ and $I_2$ respectively,
we have
\beq
I_1 \sim \int_{q_0 r}^{y_0} \frac{dy}{y} \bigl(\ln \frac{y}{r} 
    \bigr)^\beta
    \mathop{\sim}_{r \to 0} \bigl( \ln \frac{1}{r} \bigr) ^{\beta+1}
\eeq
and
\beq
I_2 \sim \int_{y_0}^{\infty} \frac{dy}{y} J_0(y) \bigl( \ln y + \ln 
    \frac{1}{r} \bigr)^{\beta}  \mathop{\sim}_{r \to 0}
     \bigl( \ln \frac{1}{r} \bigr) ^{\beta}
\eeq
So the asymptotic small $r$ behaviour will be divergent, precisely
\beq
\psi(r)  \mathop{\sim}_{r \to 0} \bigl( \ln \frac{1}{r} \bigr) ^{\beta+1}
\eeq
\section {Acknowledgments.}
The authors express their deep gratitude to Prof. C.Pajares for
their kind attention and
helpful discussions.  
M.A.B. thanks the INFN for financial help during his stay
at Bologna University and IBERDROLA for financial support during his stay at
the University of Santiago de Compostela. G.P.V. thanks 
Prof. L.Miramontes for hospitality during his stay at the University of
Santiago de Compostela.


\section{Figure captions}

\noi Fig.1. The substitution for the Green function
necessary to calculate the inclusive  jet production cross-section.

\noi Fig.2. Inclusive jet production cross-sections for the process
 $\gamma^*\gamma^*$ at  $\sqrt{s}=540GeV$ with $Q^2=25(GeV/c)^2$
and $P^2=1(GeV/c)^2$.

\noi Fig.3. Inclusive jet production cross-sections for the process
$\gamma^*p$ at $\sqrt{s}=540GeV$ with $Q^2=25(GeV/c)^2$. 

\noi Fig.4. Inclusive jet production cross-sections for the process $pp$
at $\sqrt{s}=540GeV$.

\noi Fig.5. Cross-sections $d\sigma/dy$ for the $pp$ process
at $\sqrt{s}=540GeV$.

\noi Fig.6. Multiplicities $\langle n \rangle$ as a function of the center
of mass  energy $\sqrt{s}$ for the processes $\gamma^*\gamma^*$ 
(the solid curve),$\gamma^*p$ 
       (the dashed curve) and $pp$  (the dotted curve).

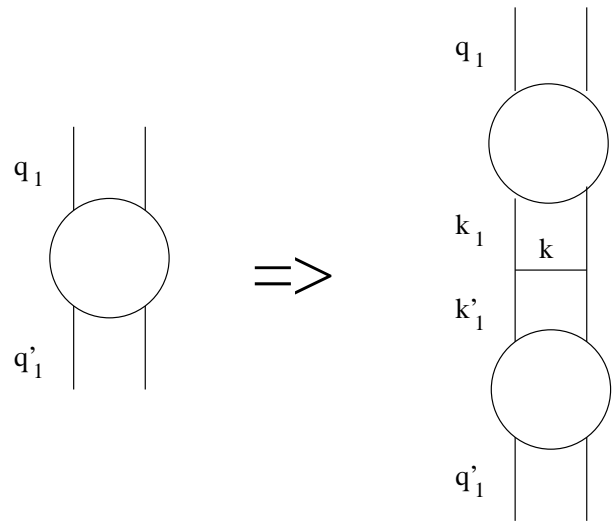
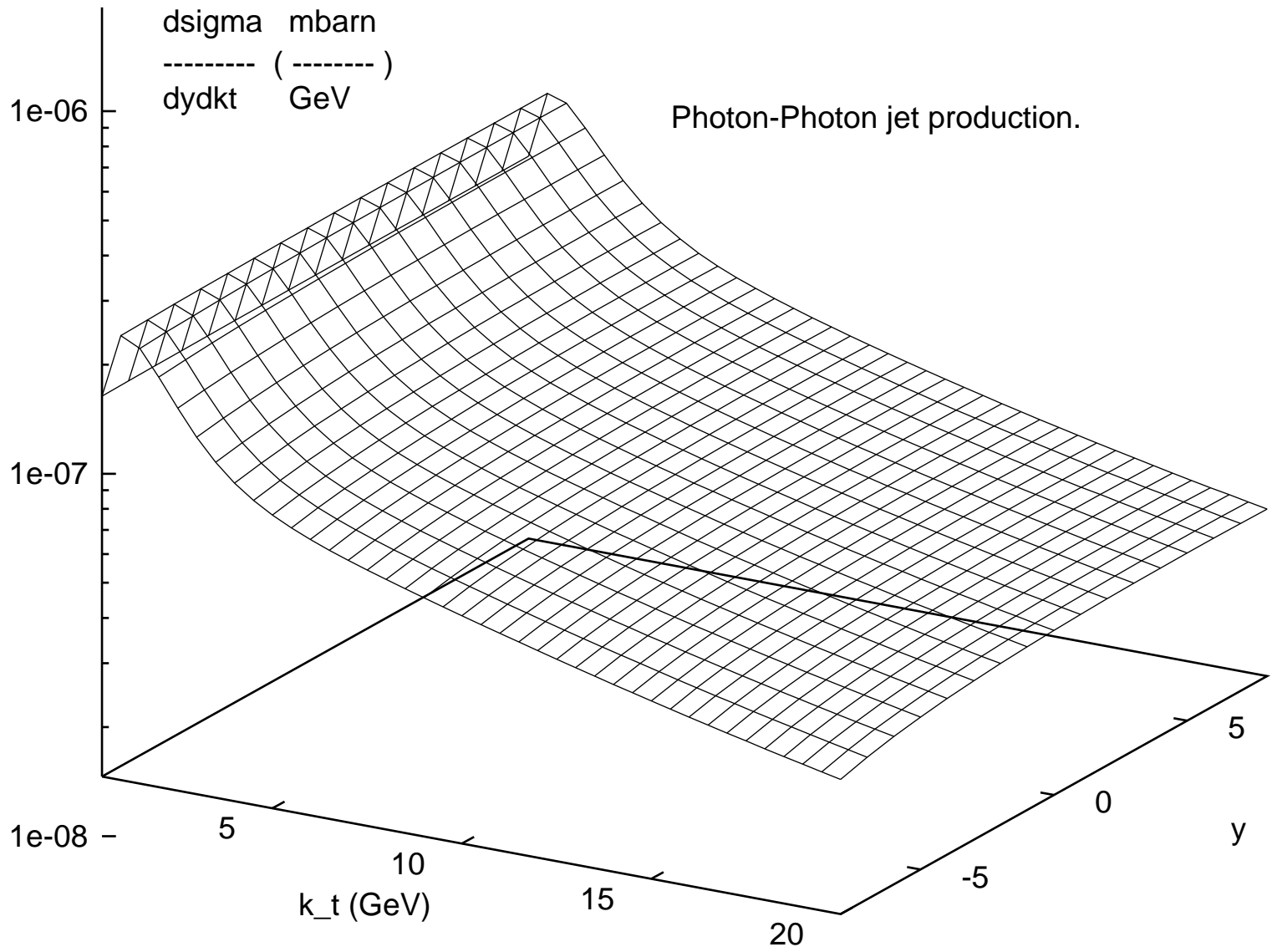
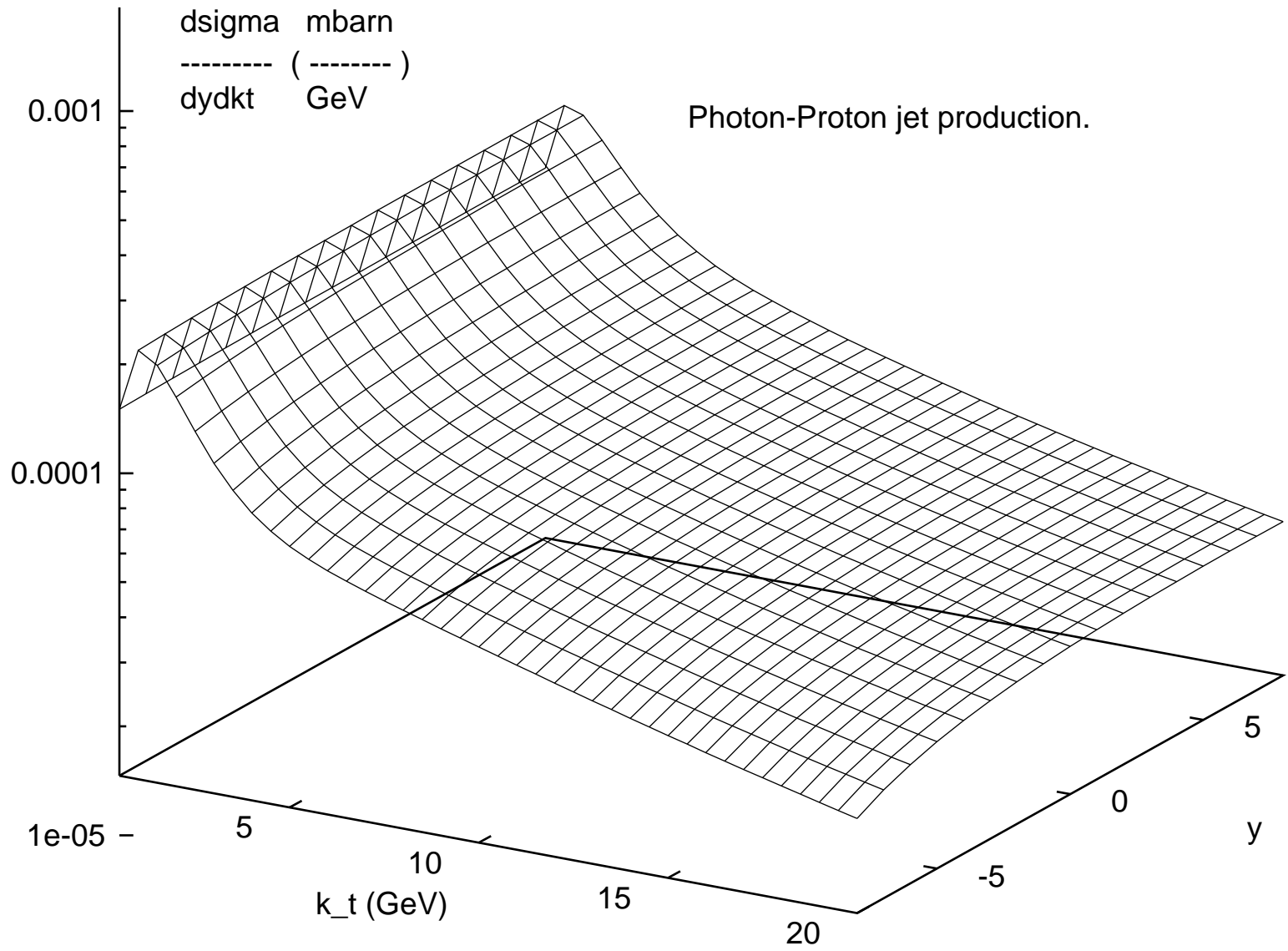
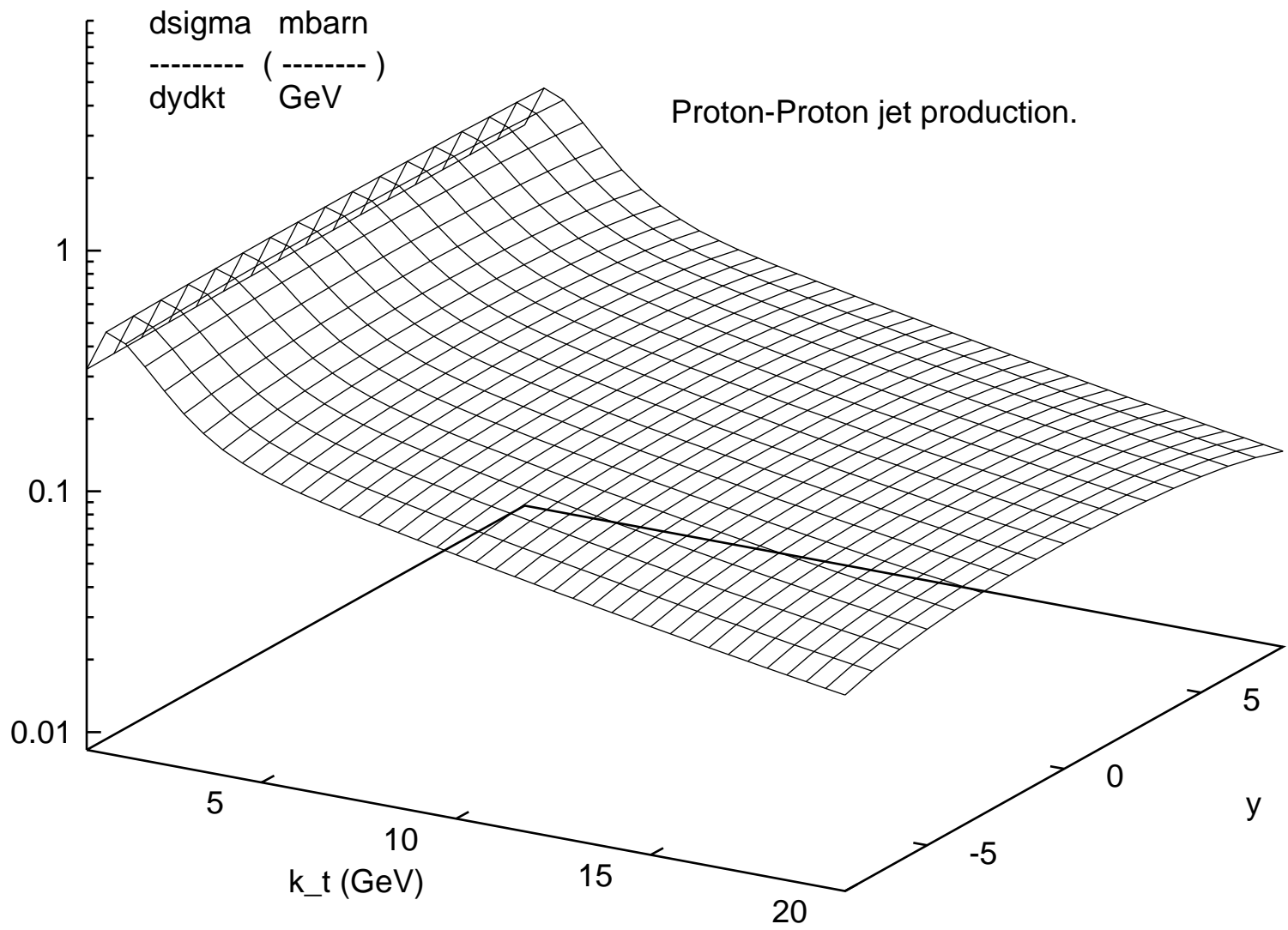
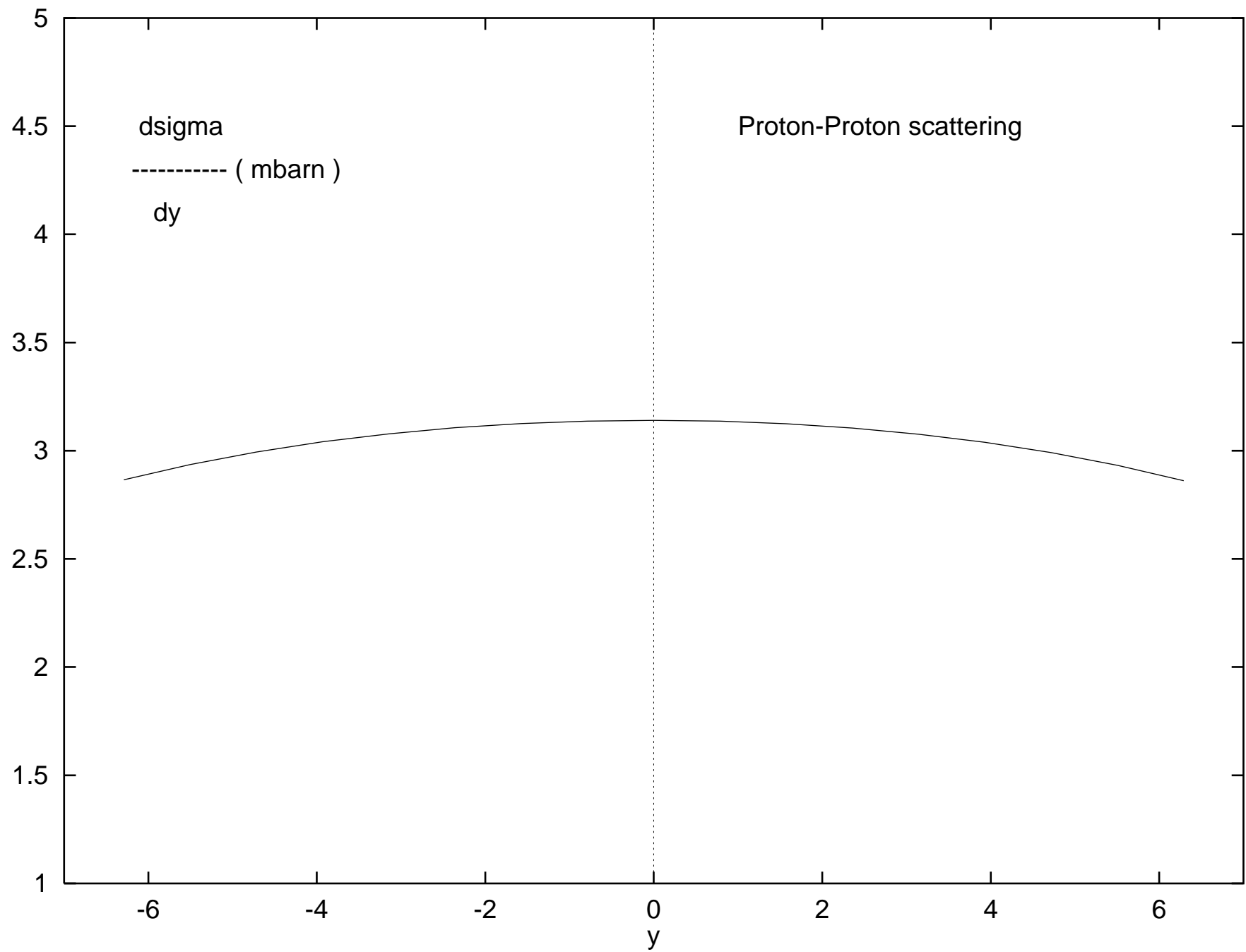
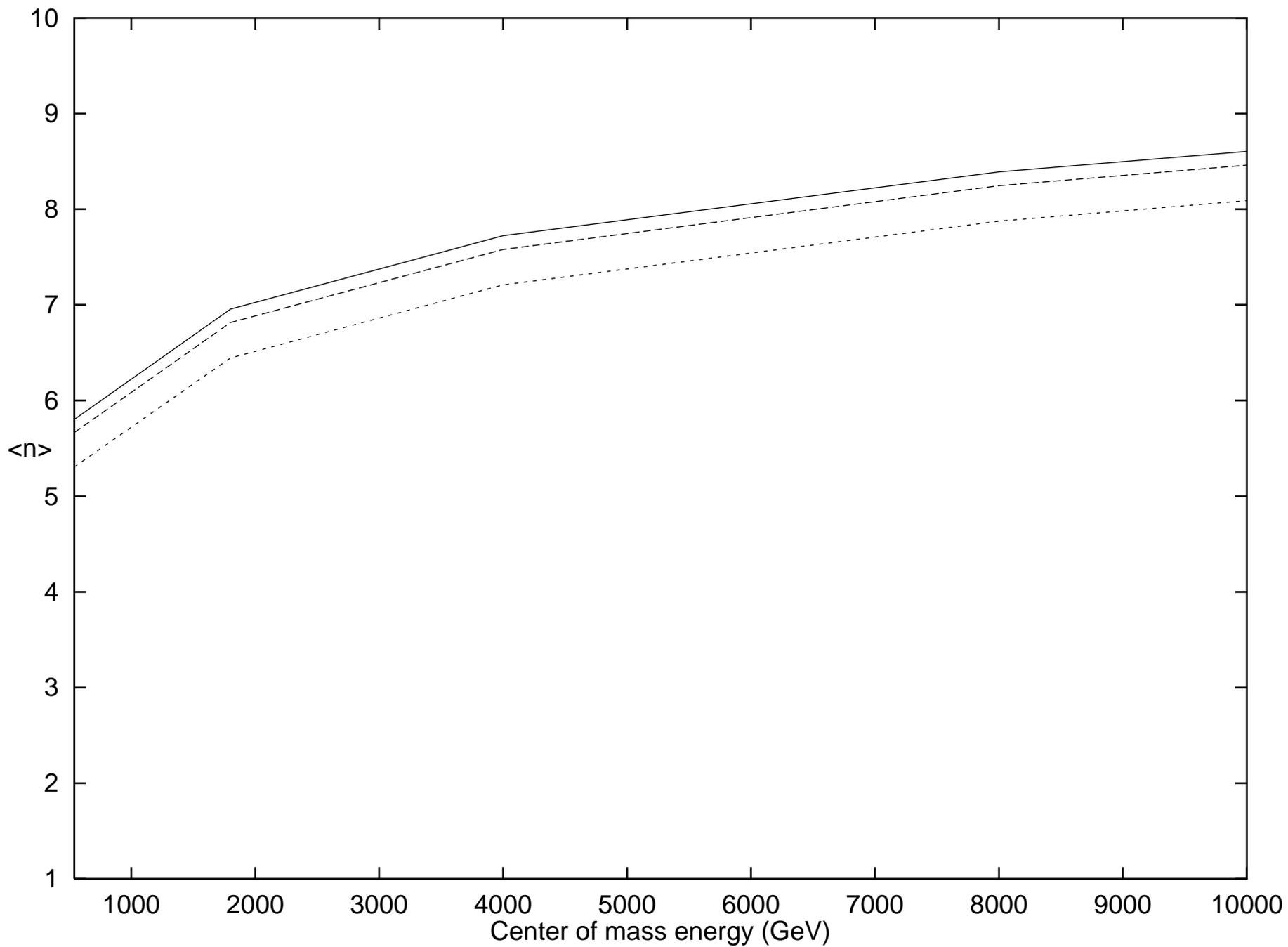
\end{document}